\documentstyle[aps,amsfonts,prbbib]{revtex}
\draft

\def\vs{\vec{s}}
\def\ce{{\cal E}}

\def\Th{\Theta}
\newcommand{\ket}[1]{\left|#1\right>}

\begin{document}
\title{Combinatorics, wreath products, finite space groups and 
  magnetism}
\author{Wojciech FLOREK, Stanis{\l}aw WA{\L}CERZ}
\address{Institute of Physics, Adam Mickiewicz University, 
ul. Umultowska 85, 61--614 Pozna\'n, Poland}
\date{January 10, 1998}
\maketitle

\begin{abstract}
In this lecture some mathematical tools necessary for a proper 
description of the Heisenberg antiferromagnet are presented. We would like 
to point out differences between ferro- and antiferromagnetic cases of 
Heisenberg Hamiltonian for finite spin systems. The ground--state properties 
are discussed. 
\end{abstract}

\section{Introduction}
The Heisenberg model of magnetism has been investigated for years. 
For the finite spin system consisting of $N<\infty$ 
spins $\vec{s}$ we obtain the following formula
\begin{equation}
{\cal H} = -J\sum_{\left<ij\right>}\:\vs_i\vs_j - h\sum_i\:s_i^z
\end{equation}
where the first sum is taken over all nearest-neighbor pairs $\left<ij\right>$ 
and $h$ is an external magnetic field parallel to $z$-axis (number of the 
nearest-neighbor pairs will be denoted hereafter as ${\cal N}$). 

\subsection{Ferromagnet ($J>0$)}
The ground state of the finite Heisenberg ferromagnet has the following 
properties:
\begin{itemize}
\item for $h=0$ the ground state is $(2Ns+1)$-tuplet with total spin
  number $S=Ns$ and energy per spin $\ce_0^+=-J{\cal N}s^2/N$;
\item For $h>0$ the above multiplet splits into singlets and the ground-state 
  is the one of them with maximal magnetization $M=Ns$ and the energy per site
  $\ce_h^+=\ce_0^+-hs$;
\item the energy gap $\Delta\ce^+$ between the ground state and the first 
   excited state is equal to $h/N$, so it is proportional to $h$.
\end{itemize}

The above properties agree with the classical results (all spins are 
`parallel') 
and with the thermodynamic limit ($N\longrightarrow\infty,\; h\longrightarrow0$).

\subsection{Antiferromagnet ($J<0$)}
The classical ground state of antiferromagnet is described by the 
so-called N\'eel 
state configuration $\ket{s\,-s\,s\ldots s\,-s}$. It is evident that this 
state has two-fold degeneracy. If one defines (see \cite{iske})
the N\'eel state as a state with opposite magnetizations in (ferromagnetic)
sublattices (i.e.\ $S_A=S_B=Ns/2$ and $M_A=-M_B$), then the degeneracy 
of this state is $Ns+1$. On the other hand Marshall and Peirels have 
proved that the ground state of antiferromagnet is a singlet with
$S=M=0$ \cite{mar}. Moreover, simple calculations for finite spin systems 
show that the ground state has the following properties
\begin{itemize}
\item the ground state is a singlet even for $h=0$;
\item the ground state is a linear combination of all states with total
      magnetization $M=0$;
\item the energy gap $\Delta\ce^-$ decreases for increasing $h>0$ and 
      for sufficient large $h$ the ground state is a state with
      $S=M=1$ (for very large $h$ in the ground state $S=M=Ns$ --- 
      the ferromagnetic ground state is obtained).
\end{itemize}

\subsection{Finite Lattice Method}
Considerations of finite spin systems is very popular and effective
method, therefore it has been frequently applied since the pioneer
work of Bonner and Fisher \cite{bf}. Number of states, which should be 
considered, is $(2s+1)^N$, so it grows very quickly (for $s=1/2$ and 
$4\times4$ square lattice there are $2^{16}=65536$ states!). Therefore a lot 
of methods are used in order to decrease a dimension of the Hamiltonian 
eigenproblem or to simplify the solving procedure (e.g.\ Lanczos method 
\cite{mor}, combinatorial methods \cite{iske} and group-theoretical 
method \cite{dux}). In the last case a translation group of considered
lattice is taken into account, as a rule. In contrary, we investigate 
also a point group and a space group (of finite lattice), therefore 
a more complete state classification scheme can be obtained \cite{lul}.  

\subsubsection{Example: Four spins $s=1/2$ with the periodic boundary
conditions}
In this case the translation group is ${\cal T}=C_4$, the point group
is ${\cal P} = \{E,\sigma\} = C_s$ and the space group --- ${\cal S}=C_{4v}$.
Classification scheme for $2^4=16$ states is given in Tab.\ref{t1}. 
$\ce$ denotes the energy per spin, $S,M$ --- total spin number and 
magnetization, respectively, and 
$\Theta\in\{\Theta_{-1},\Theta_0,\Theta_1,\Theta_2\}$,
$\Xi\in\{\Xi_0,\Xi_1\}$ and $\Gamma\in\{A_1,A_2,B_1,B_2,E\}$ are the 
irreps of $C_4$, $C_s$ and $C_{4v}$, respectively. It is worth noting 
that two possible decompositions of the irrep 
$E = \Theta_{-1}\oplus\Theta_1 = \Xi_0\oplus\Xi_1$ correspond to two 
choices of the basis in the irreducible subspace labelled by $E$. In the 
first case the basis is complex and in the second --- real one, respectively.
In other words all states labelled by $E$   correspond to wave vector 
with $|\vec{k}|=1$ but in the first case there is additional index ---
$\mbox{sign}k=\pm1$, and in the second --- a symmetry index $\alpha=\pm1$
(i.e.\ $\sigma\ket{E\alpha}=\alpha\ket{E\alpha}$, where $\sigma\in{\cal P}$
is a reflection). The (antiferromagnetic) ground state is given as a linear
combination
\begin{equation}
\frac{\sqrt{3}}{3}\left(\ket{+-+-}+\ket{-+-+}\right)-
\frac{\sqrt{3}}{6}\left(\ket{++--}+\ket{+--+}+\ket{--++}+\ket{-++-}\right).
\end{equation}

\begin{table}
\caption{Classification of states for 4 spins 1/2, $J=-1$}
\begin{center}
\begin{tabular}{||c|c|c||c|c|c||c|c||} \hline\hline
$\Th$ & $\Gamma$ & $\Xi$ & $S$ & $M$ & $\cal E$ & 
\multicolumn{2}{c||}{Degeneracy}\\ \cline{7-8}
      &          &       &     &     &          &  $h=0$ & $h\neq0$\\ \hline 
\hline                      
$\Th_0$ & $A_1$  & $\Xi_0$ & $0$ & $0$ & $-0.50$ & 1 & 1\\    
$\Th_2$ & $B_1$  & $\Xi_0$ & $1$ & $0$ & $-0.25$ & 1 & 1\\    
$\Th_2$ & $B_1$  & $\Xi_0$ & $1$ & $\pm1$ & $-0.25(1\pm h)$ & 2 & 1+1\\    
$\Th_2$ & $B_2$  & $\Xi_1$ & $0$ & $0$ & $0.00$ & 1 & 1\\    
$\Th_{-1}\oplus\Th_1$ & $E$  & $\Xi_0\oplus\Xi_1$ & $1$ & $0$ & $0.00$ & 2 & 
2\\    
$\Th_{-1}\oplus\Th_1$ & $E$  & $\Xi_0\oplus\Xi_1$ & $1$ & $\pm1$ & $\mp0.25h$ 
& 4 & 2+2\\    
$\Th_0$ & $A_1$  & $\Xi_0$ & $2$ & $0$ & $0.25$ & 1 & 1\\    
$\Th_0$ & $A_1$  & $\Xi_0$ & $2$ & $\pm1$ & $0.25(1\mp h)$ & 2 & 1+1\\    
$\Th_0$ & $A_1$  & $\Xi_0$ & $2$ & $\pm2$ & $0.25(1\mp 2h)$ & 2 & 1+1\\  
\hline\hline  
\end{tabular}
\end{center}
\label{t1}
\end{table}

\section{Method}
\subsection{Short Description}
The most important aim of our work is to determine the ground state of 
(finite) Heisenberg antiferromagnet and its properties. It has been
done for spin systems with $s=1/2$ and a linear chain up to 16 spins, square 
$4\!\times\!4$ lattice, and a $2\!\times\!2\!\times\!2$ cube. The results 
and the detail description is presented elsewhere (see \cite{varia}). The main 

points of used procedure are following  
\begin{enumerate}
\item Find number $N_0$ of states with total magnetization $M=0$ (more 
precisely --- 
      we calculate a dimension of subspace $L_0$ containing such states);
\item Determine this states, i.e.\  determine the basis $\cal B$ in the 
      subspace $L_0$;
\item Decompose this basis into orbits of the space group (since 
      the space group $\cal S$ is a subgroup of the symmetry group
      $\Sigma_N$, then one can consider the action of 
      $\Sigma_N$ on $\cal B$);
\item It can be proved that when $Ns$ is even number then the ground state
      is ``fully'' symmetric (i.e.\ it transforms as the unit irrep), 
therefore
      from each subspace spanned on a given orbit one (the unique) such state 
      is chosen;
\item The eigenproblem for the operator $\vec{S}^2$ (square of total spin 
      $\vec{S}$) is solved for these states (the eigenvalues of this 
      operator are $0,2,6,\ldots,Ns(Ns+1)$ and the equation
      $\vec{S}^2\ket{\psi}=0$ is the most interesting);
\item After the above presented steps the states labelled by $M=0$, 
      $\Gamma=\Gamma_0$, and $S=0$ are obtained and the ground state is a 
      linear combination of these states --- it is determined by 
      solution of the eigenproblem ${\cal H}\ket{\varphi}=\ce\ket{\varphi}$;
\item As a result the ground state is obtained as a linear combination
      of the so-called Ising configurations and its properties can be 
      easily determined (e.g.\ spin-spin correlations, staggered 
magnetization,
      etc.).     
\end{enumerate} 
It should be underlined that this procedure gives only the ground state, 
therefore
the thermodynamics properties of the considered system cannot be determined. 
These properties can be found when one solves eigenproblems for each
total spin number $S=0,1,2,\ldots,Ns$ and for each irrep of the space group.

\subsection{Combinatorics}
The first three steps can be done applying combinatorial methods. The problem 
is: `Find all states with a given magnetization, i.e.\ states 
$\ket{m_1m_2\ldots m_N}$ fulfilling the condition $\sum_i\;m_i=M$'. Since
the magnetization operator $S^z=\sum_i\;s_i^z$ commutes with any 
$\sigma \in \Sigma_N$, then the action of $\Sigma_N$ on the set 
$\{1,2,\ldots,N\}$ can be considered. In the simplest case $s=1/2$ number 
of states with the magnetization equal $M$ is given by the binomial 
coefficient
$$
\dim L_M = \left(\begin{array}{c}N\\ \frac{N}{2}+M\end{array}\right)
$$
where $k=N/2+M$ is a number of spins with a projection $m=1/2$. For $s>1/2$
it can be generalized by the polynomial coefficient \cite{AM}
$$
\left( \begin{array}{c} N\\n_0n_1\ldots n_{2s}\end{array}\right)
=\frac{N!}{n_0!n_1!\ldots n_{2s}!};\;\;\;\sum_{i=0}^{2s}\;n_i=N
$$
where $n_i$ denotes number of spins with projection $i-s$. Number of states 
is determined by the following sum
$$
\dim L_M=\sum_{(n_0n_1\ldots n_{2s})}\left( \begin{array}{c} 
     N\\n_0n_1\ldots n_{2s}\end{array}\right)
$$
taking over all decompositions $(n_0n_1\ldots n_{2s})$ with the condition
$\sum_i\:n_i(i-s)=M$. For example, for $N=3$, $s=1$, and $M=0$ one obtains
$$
\dim L_0=
\left( \begin{array}{c} 3\\111\end{array}\right)+
\left( \begin{array}{c} 3\\030\end{array}\right)=7.
$$

\subsection{Finite space groups}
The fourth step of our procedure is to determine symmetry adapted basis of 
subspace $L_0$ according to the symmetry group of the considered Hamiltonian.
It means, that only these permutations $\sigma\in\Sigma_N$ are taking into
account which preserve order of points (`neighborhood). These elements
form (finite) space group of finite lattice. It can be shown that in the 
one-dimensional case it is the group $C_{Nv}=C_N\Box C_s=D_N$. From it follows
that for a hypercubic lattice in $d$-dimensional space the space group 
is given as a wreath product \cite{sspcm}
$$
{\cal S}=C_{Nv}\wr\Sigma_d
$$ 
where elements of the symmetric group $\Sigma_d$ permute axes of 
$d$-dimensional coordinate system. The above group is called a 
{\it complete monomial group (of degree $d$) of the group $C_{Nv}$}. 
The considerations of a linear representations of this group in the 
subspace $L_0$ give us the appropriate symmetry adapted basis, which is
used in the next steps of our procedure (5--7). These steps are performed 
using
numerical methods (solution of eigenproblem for real symmetric matrix). 

\section{Final Remarks}
Using the above described method we have obtained, e.g., energy per site
for 16-spin linear chain. Its value --- $-0.446$ --- is very close to the exact
result in the thermodynamic limit --- $1/4-\ln 2 \approx -0.4432$. Therefore,
one can say that in the one-dimensional case a system of 20 spins is
quite good approximation of the infinite system.

Application of the proposed procedure to two- and three-dimensional spin
systems requires much more intensive investigations of group action of
the symmetric group $\Sigma_N$ on the basis states (the Ising configurations)
$\ket{m_1\ldots m_N}$. It should enable to consider system of 20$\times$20 
or $10\times10\times10$ spins for 
any value of spin number $s$.

\end{document}